\documentclass{article}

\usepackage{arxiv}

\usepackage[utf8]{inputenc} 
\usepackage[T1]{fontenc}    
\usepackage{hyperref}       
\usepackage{url}            
\usepackage{booktabs}       
\usepackage{amsfonts}       
\usepackage{nicefrac}       
\usepackage{microtype}      

\usepackage{graphicx}

\usepackage{colortbl}		
\usepackage{multicol}        

\usepackage{amsbsy}
\usepackage{amsfonts}
\usepackage{amsmath}

\usepackage{amsfonts}
\usepackage{amsmath}
\usepackage{xcolor}
\usepackage{booktabs} 

\newcommand{\bs}[1]{\mathbf{#1}}
\newcommand{\x}{\bs x(\bs s)}
\newcommand{\y}{\bs y(\bs s)}
\newcommand{\z}{\bs z(\bs s)}
\newcommand{\m}{\bs m}
\newcommand{\s}{\bs s}
\newcommand{\A}{\bs A}
\newcommand{\W}{\bs W}
\newcommand{\M}{\bs M}
\newcommand{\U}{\bs U}
\newcommand{\V}{\bs V}

\newcommand{\LCov}{\text{LCov}}
\newcommand{\LDiff}{\text{LDiff}}
\def\Cov{\mathop{\mathrm{Cov}}\nolimits}

\def\E{\mathop{\mathrm{E}}\nolimits}
\newcommand{\dom}{\mathcal{S}}

\newcommand{\R}{\mathbb{R}}
\newcommand{\F}{\mathcal{F}}

\newtheorem{definition}{Definition}

\title{Spatial Blind Source Separation in the Presence of a Drift}

\date{} 					

\author{Christoph~Muehlmann \\
	Institute of Statistics \& Mathematical Methods in Economics \\
	Vienna University of Technology, Austria \\
	\texttt{christoph.muehlmann@tuwien.ac.at} \\
	\And
	Peter Filzmoser\\
	Institute of Statistics \& Mathematical Methods in Economics \\
	Vienna University of Technology, Austria \\
    \texttt{peter.filzmoser@tuwien.ac.at}
	\And
	Klaus~Nordhausen \\
	Department of Mathematics and Statistics \\
	University of Jyv\"askyl\"a, Finland \\
	\texttt{klaus.k.nordhausen@jyu.fi} \\}

\renewcommand{\headeright}{}
\renewcommand{\undertitle}{}

\begin{document}
\maketitle

\begin{abstract}
Multivariate measurements taken at different spatial locations occur frequently in practice. Proper analysis of such data needs to consider not only dependencies on-sight but also dependencies in and in-between variables as a function of spatial separation. Spatial Blind Source Separation (SBSS) is a recently developed unsupervised statistical tool that deals with such data by assuming that the observable data is formed by a linear latent variable model. In SBSS the latent variable is assumed to be constituted by weakly stationary random fields which are uncorrelated. Such a model is appealing as further analysis can be carried out on the marginal distributions of the latent variables, interpretations are straightforward as the model is assumed to be linear, and not all components of the latent field might be of interest which acts as a form of dimension reduction. The weakly stationarity assumption of SBSS implies that the mean of the data is constant for all sample locations, which might be too restricting in practical applications. Therefore, an adaptation of SBSS that uses scatter matrices based on differences was recently suggested in the literature. In our contribution we formalize these ideas, suggest an adapted SBSS method and show its usefulness on synthetic and real data.
\end{abstract}

\section{Introduction}

Data that are collected at different spatial locations occur quite often in statistical modeling. Pollution measurements in cities, element concentration in mines or socioeconomic variables across countries are examples of measurements which show dependencies as a function of spatial separation. With advanced technology data collection is significantly improved which leads to higher numbers of sample locations as well as measured variables. Therefore, proper statistical tools for such multivariate spatial data need to consider spatial dependencies in and in-between variables of interest.

In a first exploratory step, multivariate spatial data might be analyzed by using the popular Principal Component Analysis (PCA), where variance maximizing orthogonal transformations of the data are found. The disadvantage of this method in the present context is obvious. Only variance is maximized which ignores the spatial second order dependencies completely, orthogonal transformations might be too restricting, and the result is dependent on the scale of the data. \cite{nordhausen2015blind,bachoc2018spatial} introduced Spatial Blind Source Separation (SBSS) as an adaptation of Blind Source Separation (BSS) for spatial data, which overcomes the significant drawbacks of PCA. BSS is a framework which originates from signal processing and assumes that the multivariate data at hand are formed by linear transformations (not necessary orthogonal) of unobserved variables. The aim of BSS is to estimate these latent variables which might show a clearer structure and reveal the driving processes of the data. BSS is well-established for many types of data, such as independent and identically distributed (iid) data, where it is denoted as Independent Component Analysis (ICA), \cite{NordhausenOja2018}  time series data \cite{PanMatilainenTaskinenNordhausen2021} or tensorial data \cite{VirtaLiNordhausenOja2017}. For general overviews see also \cite{CichockiAmari2002,comon2010handbook}. SBSS, in detail described in Section~\ref{sec:sbss}, assumes that the observed multivariate data are formed by (spatially) uncorrelated, weakly stationary latent random fields. The motivation behind this assumptions is clear. Firstly, as the entries of the latent random field are uncorrelated, univariate analysis can be carried out individually which discards demanding multivariate approaches, \cite{kriging_paper}. Secondly, the latent random field is found by maximizing second order spatial dependence but interpretations are still straightforward as the identified transformations are linear. Lastly, only certain components might be of interest for further analysis or the domain expert which acts as a way of dimension reduction. SBSS finds the latent random field by jointly diagonalizing the covariance and so-called local covariance matrices that capture second-order spatial dependence.

The original assumption of SBSS that the entries of the latent field are second order stationary might be too restricting in practical considerations. In many situations it is more natural to consider a drift in the data which violates the non-constant mean assumption. \cite{CAJG_paper} considered that case by replacing local covariance matrices with local difference matrices, which avoids the estimation of the mean and is practically more robust in the presence of a drift. Section~\ref{sec:ldif} is devoted to the use of local difference matrices in SBSS, puts the ideas of \cite{CAJG_paper} on a solid basis and introduces a new SBSS method with an adapted whitening step. The usefulness of this new method is validated on synthetic datasets in Section~\ref{sec:sim} and illustrated on a geochemical dataset in Section~\ref{sec:real_data}. Section~\ref{sec:conclusion} presents concluding remarks and an outlook for upcoming research.

\section{Spatial Blind Source Separation}\label{sec:sbss}

BSS for spatial data is a relatively new field in geostatistics. It was first introduced by \cite{nordhausen2015blind} where the method was motivated by a geochemical application. \cite{bachoc2018spatial} put SBSS on a sound theoretical basis, refined SBSS, and derived asymptotic properties for the estimators. Both publications are based on the following statistical model.

\begin{definition}[SBSS model]\label{def:sbss_model}
A $p$-variate random field $\x$ defined on a $d$-dimensional spatial domain $\dom \subseteq \R ^ d$ follows a spatial blind source separation model if it can be written as
\begin{equation}\label{eq:sbss}
	\x = \A \z + \bs m,
\end{equation}
for all $\s \in \dom$, where $\A$ is the full-rank $p \times p$ deterministic mixing matrix, $\bs m$ is the constant, $d$-dimensional, deterministic drift vector and $\z$ is the $p$-variate latent random field which fulfills the following assumptions.
\begin{description}
 \item[(SBSS 1):] $\E (\z ) = \bs{0}$ for all $\s \in \dom$,
 \item[(SBSS 2):] $\Cov (\z ) = \E \left(\z \z ^ \top \right) = \bs I_p$ for all $\s  \in \dom$ and
 \item[(SBSS 3):] $\Cov (\bs{z}(\bs{s}),  \bs{z}(\bs{s}') ) = \E (\bs{z}(\bs{s}),  \bs{z}(\bs{s'}) ^ \top) = \bs D(\bs{s} -  \bs{s}')$ for all $\s , \s' \in \dom$ with $\s \neq \s'$, where $\bs D$ is a diagonal matrix containing the stationary covariance functions of each entry of $\z$ as diagonal elements.
\end{description}
\end{definition}

Note that the SBSS model is a semi-parametric linear latent variable model as $\z$ is unobserved and only assumptions on the first two moments are stated. Specifically, the latent field $\z$ is constituted by uncorrelated weakly stationary univariate random fields. Furthermore, $\z$ is only defined up to sign and order as $\bs P \bs S \z$ still fulfills all assumptions stated above, here $\bs P$ is a permutation matrix and $\bs S$ is a diagonal matrix where each diagonal element is either $+1$ or $-1$. In this context this is of minor importance as the sign and order might be determined by the context of the analysis or is not of interest at all. Note also that the scale is not identifiable for BSS in general, but in the model stated above it is fixed by assumption (SBSS 2) to be unity; this assumption will be given up in a subsequent section.

The aim of SBSS is to recover $\z = \W (\x - \m)$ up to sign and permutations only based on one given realization of $\x$ on $n$ sample locations by estimating the so-called unmixing matrix $\W$ and the drift $\m$. Generally, the estimation of $\W$ is usually carried out in a two-step procedure by almost all BSS methods as follows. The singular value decomposition of the mixing matrix yields $\A = \U \bs D \V ^ \top$ which determines the covariance matrix of $\x$ to be $\Cov(\x) = \U \bs D ^ 2 \U ^ \top$. As the covariance matrix is positive definite by assumption one can whiten $\x$ by $\Cov^{-1/2}(\x)(\x - \m)$ which equals $\Cov^{-1/2}(\x) \A \z = \U \V ^ \top \z$ by plugging in Equation~\eqref{eq:sbss}. This suggests to firstly whiten $\x$ and then to find only an orthogonal matrix to recover $\z$. More details and a proper mathematical derivation can be found in \cite{miettinen2015}.

In order to find the orthogonal matrix after the whitening step, \cite{nordhausen2015blind} and \cite{bachoc2018spatial} introduced local covariance matrices ($\LCov$) as a key tool for SBSS, which are defined as
\[
\LCov_f(\x) = \frac{1}{n} \sum_{i,j = 1}^n f(\s_i - \s_j) \E \Bigl[ [\bs x(\s_i) - \E (\bs x(\s_i))] [\bs x(\s_j)- \bs \E (\bs x(\s_j)) ]^\top \Bigr].
\]

$\LCov$ matrices are weighted averages of the covariance matrices between all possible pairs of $n$ given sample locations, where the weights are determined by the spatial kernel function $f:\R ^ d \rightarrow \R$. \cite{nordhausen2015blind} suggested ball kernel functions $f_b(\bs h) = I(\| \bs h \| \leq r)$ which only consider locations that are separated by a maximum distance of $r$. \cite{bachoc2018spatial} additionally considered ring and Gauss kernel functions. The ring kernel function is written as $f_r(\bs h) = I(r_i < \| \bs h \| \leq r_o)$, which considers locations that are separated by a minimum of $r_i$ and a maximum of $r_o$. A smooth version of the ball kernel function is the Gauss kernel function $f_g(\bs h) = \exp(-0.5 (\Phi^{-1}(0.95) \| \bs h \| / r)^2)$ where $\Phi^{-1}(0.95)$ is the $95\%$ quantile of a standard Normal distribution. Generally, the kernel function can be of different shapes where for examples anisotropies can be modeled by accounting also for the direction of $\bs h$. $\LCov$ matrices are symmetric if the kernel function is symmetric, i.e. $f(\bs h) = f( - \bs h)$. Note that when $f(\bs h) = I(\| \bs h \| = 0)$, $\LCov$ matrices reduce to the usual covariance matrix, in the following denoted as $\LCov_0$.

Considering the SBSS model above, the $\LCov$ matrices evaluated on $\z$ yield a diagonal matrix which motivates the following method seen in \cite{bachoc2018spatial}.

\begin{definition}\label{def:sbss_sd}
For a random field $\x$ following the SBSS model in Definition~\ref{def:sbss_model} the unmixing matrix functional $\W$ simultaneously diagonalizes the covariance matrix and one local covariance matrix for a given kernel function $f$ such that 
\[
	\W \emph{\LCov}_0(\x) \W^\top = \bs I_p ~ \text{and} ~ \W \emph{\LCov}_f(\x) \W^\top = \bs D_f,
\]
where $\bs D_f$ is a diagonal matrix with decreasingly ordered diagonal elements.
\end{definition}

For the above method the exact diagonalizer $\W$ can be found by solving the generalized eigenproblem, or equivalently in a two step fashion by firstly whitening the data as $\bs x^{wt}(\s) = \LCov_0^{-1/2}(\x)(\x - \bs m)$ and then performing an eigendecomposition of $\LCov_f(\bs x^{wt}(\s))$. Note that the ordering of the diagonal elements of $\bs D_f$ fixes the order of the components of the latent field. As only one specific $\LCov$ matrix is utilized this method is very sensitive to its specific spatial kernel function choice. One workaround is suggested by \cite{bachoc2018spatial} leading to the following definition.

\begin{definition}\label{def:sbss_jd}
Consider a random field $\x$ following the SBSS model in Definition~\ref{def:sbss_model}. The whitened version of $\x$ is defined by $\bs x^{wt}(\s) = \emph{\LCov}_0^{-1/2}(\x)(\x - \bs m)$. For a given set of kernel functions $\F = \{ f_1,\dots, f_K \}$, $\U$ is the $p \times p$ orthogonal joint diagonalization matrix which maximizes
\[
\sum_{k=1}^K \| \text{\emph{diag}}(\U \emph{\LCov}_{f_k}(\bs x^{wt}(\s)) \U^\top ) \|^2_F.
\] 
Then, the unmixing matrix functional is given by $\W = \U^\top  \emph{\LCov}_0^{-1/2}(\x)$.
\end{definition}

In Definition~\ref{def:sbss_jd}, $\text{diag}(\bs M)$ denotes the diagonal matrix where the diagonal elements are the ones of the matrix $\bs M$, and $\| \cdot \|_F$ denotes the Frobenius matrix norm. For a given realization of $\x$ on $n$ sample locations the $K$ $\LCov_{f_k}(\bs x^{wt}(\s))$ matrices usually do not commute, therefore, algorithms that approximately jointly diagonalize these matrices have to be used. We use \cite{CardosoSouloumiac1996} which is based on iterative applications of Givens rotations. In contrast to Definition~\ref{def:sbss_sd} the method of Definition~\ref{def:sbss_jd} is more robust to the choice of spatial kernel functions but relies on a more sophisticated joint diagonalization algorithm.

The above two methods are formulated for the case of the SBSS model where the drift is assumed to be constant for all sample locations. Practically, when considering data with a present non-constant drift function, the performance of the SBSS methods above can be heavily downgraded when the population moments are replaced by their sample counterparts, especially when considering that $\E(\bs x(\s_i))$ is estimated by the sample mean for all $i=1,\dots,n$. This effect might be reduced by relying on differences, which is usually done in geostatistics where the variogram is favored over the covariance matrix. We will adapt this idea for SBSS in the following. 

\section{Local Difference Matrices for SBSS}\label{sec:ldif}

Relying on differences is common practice in geostatistics where the variogram is the central quantity of structural analysis, much favored over the spatial covariance matrix, see textbooks such as \cite{ChilesetAl1999}. Differences are also beneficial for iid or time series data. In the latter case often differences are used to stabilize the mean, or remove seasonality effects. A popular model that is based on differences is the Autoregressive Integrated Moving Average (ARIMA) model, in which differences are modeled as an Autoregressive Moving Average (ARMA) model, see for example \cite{ARIMA_ref}. For the iid case \cite{NordhausenTyler2015} emphasized that only scatter matrices which can be expressed in terms of differences possess the property that they are in every distributional case diagonal when the random vector is formed by statistically independent entries. The covariance matrix can be written in terms of differences. Furthermore, they showed theoretically that robust scatter matrices have this property as well when evaluated on differences, and practically that statistical methods such as independent component analysis (ICA) or observational regression greatly benefit from the use of differences. Therefore, we adapt local covariance matrices and introduce the following definition of local difference ($\LDiff$) matrices, as firstly mentioned in \cite{CAJG_paper}.
\[
\LDiff_f(\x) = \frac{1}{n} \sum_{i,j = 1}^n f(\s_i - \s_j) \E \Bigl[ [\bs x(\s_i) - \bs x(\s_j)] [\bs x(\s_i)- \bs x(\s_j) ]^\top \Bigr].
\]
Again, $f$ is the spatial kernel function which determines the locality of the weighted average of differences, and the same rules apply as for $\LCov$ matrices. Note that when $f$ is chosen to be a slightly adapted version of the ring kernel, then $\LDiff$ matrices yield the typical semivariogram. Practically, for a given realization of $\x$ on $n$ sample locations, the key difference between $\LCov$ and $\LDiff$ matrices in this context is that $\LCov$ matrices rely on the sample average for the drift estimate, whereas $\LDiff$ matrices do not. Theoretically, when the random field has a non-constant drift function $\m (\bs s)$ the bias is of the form $\m (\bs s) - \bs m(\s')$ for $\LDiff$ matrices. This bias is expected to be small when the kernel function is designed in such a way that the locations $\s$ and $\s'$ are nearby and under the assumption that the drift appears locally constant and would for example change only smoothly. Therefore, we suggest to replace the $\LCov$ matrix by a $\LDiff$ matrix in Definition~\ref{def:sbss_sd}. It would also be possible to adapt Definition~\ref{def:sbss_jd} in such a way, but as usually the set of kernel functions $\F$ is formed by spatial kernel functions with increasing parameters, the bias of $\LDiff$ matrices will be considerably higher. Therefore, we only consider the adaptation of Definition~\ref{def:sbss_sd}.

\begin{definition}\label{def:sbss_ldiff_sd_1}
For a random field $\x$ following the SBSS model in Definition~\ref{def:sbss_model} the unmixing matrix functional $\W$ simultaneously diagonalizes the covariance matrix and one local difference matrix for a given kernel function $f$ such that 
\[
	\W \emph{\LCov}_0(\x) \W^\top = \bs I_p ~ \text{and} ~ \W \emph{\LDiff}_{f}(\x) \W^\top = \bs D_f.
\]
Where $\bs D_f$ is a diagonal matrix with increasingly ordered diagonal elements.
\end{definition}

Again, $\W_f$ can be found by solving the generalized eigenproblem or by the two step algorithm discussed above. Note that in the above definition the whitening step relies on the use of $\LCov_0$ which still shows the disadvantages discussed above. Therefore, we suggest the following adaptation.

\subsection{Adaptation of the Whitening Step}

To robustify the whitening step in the presence of a drift we suggest to replace the $\LCov_0$ matrix by some $\LDiff$ matrix. A similar procedure was already formulated by \cite{OjaSirkiaEriksson2016} in the context of ICA, where the unmixing matrix is found by simultaneous diagonalization of two scatter matrices $\bs S_1$ and $\bs S_2$ where both not necessarily need to be the covariance matrix. Similarly, this idea was also developed for time series BSS in \cite{BelouchraniCichocki2000,GeorgievCichocki2001} where the whitening step is carried out by using a linear combination of symmetrized autocorrelation matrices to be robust in the presence of additive white noise. 

As stated above under Assumption (SBSS 2), $\M \A$ is an orthogonal matrix, where $\M = \LCov_0^{-1/2}(\x)$. This ensures that only an orthogonal matrix needs to be found when recovering $\z$ after the data is whitened with respect to $\LCov_0^{-1/2}(\x)$. Assumption (SBSS 2) can be replaced by assuming that $\LDiff_f(\z) = \bs I_p$ which leads to $\M \A$ being orthogonal as well when $\M = \LDiff_f^{-1/2}(\x)$. In that case the whitening step is carried out with respect to $\LDiff_f^{-1/2}(\x)$ and it is again ensured that only an orthogonal matrix needs to be found when recovering $\z$. This comes with the advantage that also in the whitening step the local covariance matrix is replaced by a local difference matrix at the cost of the fixed scale of $\z$. In a similar context this procedure is denoted as Robust Orthogonalization \cite{GeorgievCichocki2001}. This outline leads to the following definition.


\begin{definition}\label{def:sbss_ldiff_sd_2}
Consider two spatial kernel functions $f_1$ and $f_2$. For a random field $\x$ following the SBSS model in Definition~\ref{def:sbss_model} where Assumption (SBSS 2) is replaced by (SBSS $2^*)$: $\emph{\LDiff}_{f_1}(\z) = \bs I_p$. The unmixing matrix functional $\W$ simultaneously diagonalizes the corresponding two local difference matrices such that 
\[
	\W \emph{\LDiff}_{f_1}(\x) \W^\top = \bs I_p ~ \text{and} ~ \W \emph{\LDiff}_{f_2}(\x) \W^\top = \bs D_{f_1 f_2}.
\]
Where $\bs D_{f_1 f_2}$ is a diagonal matrix with increasingly ordered diagonal elements.
\end{definition}

As shown in \cite{CichockiAmari2002}, Theorem 4.2, the mixing matrix can be found when $\LDiff_{f_1}(\x)$ is symmetric positive definite and $\LDiff_{f_2}(\x)$ is symmetric and the generalized eigenvalues (diagonal elements of $\LDiff_{f_2}(\z)$) are distinct. It is easy to see that $\LDiff$ matrices are symmetric and positive definite in the context of Definition~\ref{def:sbss_model} when considering ball, ring and Gauss kernel functions with strictly positive parameters. Again as in Definition~\ref{def:sbss_sd} the unmixing matrix can be alternatively found in a two step fashion where the whitening is now carried out by $\bs x^{wt*}(\s) = \LDiff^{-1/2}_{f_1}(\x) (\x - \m)$. Note that for this adapted whitening procedure $\LCov_0(\bs x^{wt*}(\s))$ is not necessarily diagonal but $\LCov_0(\W \x)$ is diagonal but not equal to $\bs I_p$. Therefore, in this adaptation the convenience of fixing the scale of $\z$ is given up for the sake of practical more robust whitening with respect to a present drift. In practical considerations $\W \x$ might be standardized by scaling each component to unit variance. We also want to emphasize that the whitening step in Definition~\ref{def:sbss_sd} and \ref{def:sbss_jd} can be adapted in a similar way by replacing $\LCov_0$ by $\LCov_f$ for some spatial kernel function $f$, but in contrast to the above outline $\LCov_f$ is not necessarily positive definite which might be overcome by the approach described in \cite{BelouchraniCichocki2000}.

\subsection{Comments on the Drift}

For data that is generated by the model of Definition~\ref{def:sbss_model}, the latent random field can be recovered by $\z = \A (\x - \m)$ up to sign, order (and also scale when considering the context of Definition~\ref{def:sbss_ldiff_sd_2}). But often one finds in practical considerations that the data at hand shows a non-constant drift which violates the SBSS model assumption of a constant mean. In such a situation we suggest to use the methods of Definition~\ref{def:sbss_ldiff_sd_1} or \ref{def:sbss_ldiff_sd_2} for the reasons outlined above, compute $\y = \W (\x)$ and handle the present drift in one of the following forms based on the aim of further analysis of the data. All the following relies on the fact that $\bs y (\s)$ consists of uncorrelated entries, which is the great advantage of the SBSS framework.


If one is interested in predicting $\x$ at some unobserved location $\s^*$, each entry of $\bs y (\s)$ can be treated individually. Universal Kriging would be the natural choice as it is designed for the presence of a drift, but also any other prediction tool might be used, see textbooks such as \cite{ChilesetAl1999}. After predicting each entry of $\bs y (\s)$ individually, the vector of predictions $\hat{\bs y} (\s^*)$ can be formed and transformed by $\W^{-1} \hat{\bs y} (\s^*)$ to obtain a predictions of the original field $\hat{\bs x} (\s^*)$. This discards the use of multivariate prediction tools in favor of $p$ univariate ones, which reduces the complexity of modeling significantly. This procedure was already investigated and described in detail for the constant drift case in \cite{kriging_paper}.

If the aim of further analysis is to recover the latent field without any drift, each entry of the present drift $\m (\s)$ of $\y$ can be estimated individually, for example by a univariate Kriging estimator. As before the vector of predictions $\hat{\bs m} (\s)$ can be formed and subtracted from $\bs y (\s)$, which results in an practical estimation of the latent random field without drift. Additionally, the predicted drift can be back-transformed by $\W^{-1} \hat{\bs m} (\s)$ to obtain an estimation of the original drift of $\x$. As before the great advantage is that this procedure replaces the use of one multivariate model by $p$ univariate ones.

\subsection{SBSS for Intrinsic Stationary Random Fields}\label{sec:intrinsic_stat_sbss}

In the following we discuss an adaptation of the SBSS model in Definition~\ref{def:sbss_model} for which the procedure of Definition~\ref{def:sbss_ldiff_sd_2} is the natural choice. We adapt the SBSS model (Definition~\ref{def:sbss_model}) by assuming that $\z$ is formed by uncorrelated intrinsic stationary random fields. More formally this yields to replace assumptions (SBSS 2) and (SBSS 3) in the following way.
\begin{description}
 \item[(SBSS $2^*$):] $\Cov (\z ) = \E \left(\z \z ^ \top \right) = \bs D(\s)$ for all $\s  \in \dom$ where $\bs D(\s)$ is a diagonal matrix with strictly positive diagonal entries, and
 \item[(SBSS $3^*$):] $\E \left( (\z - \bs{z}(\bs{s'})) (\z - \bs{z}(\bs{s'})) ^ \top \right) = \bs K(\bs{s} -  \bs{s}')$ for all $\s , \s' \in \dom$ with $\s \neq \s'$, where $\bs K$ is diagonal holding two times the variogram for each entry of $\z$ on its diagonal.
\end{description}
For this model the estimation of $\LCov_f(\x)$ can be highly corrupted by the fact that the spatial covariance function for intrinsic stationary random fields is a function of the sample locations themselves and not of the distance between them. In contrast, the estimation of $\LDiff$ matrices is not corrupted as the variograms are still only dependent on the distance between sample locations. Therefore, the only choice for estimating the unmixing matrix which is consistent with the adapted SBSS model is the one from Definition~\ref{def:sbss_ldiff_sd_2}, something we will also confirm by the simulations in the following section.

\begin{figure}[t]
\centering
    \begin{minipage}[t]{0.42\textwidth}    
        \centering
        \includegraphics[width=\linewidth]{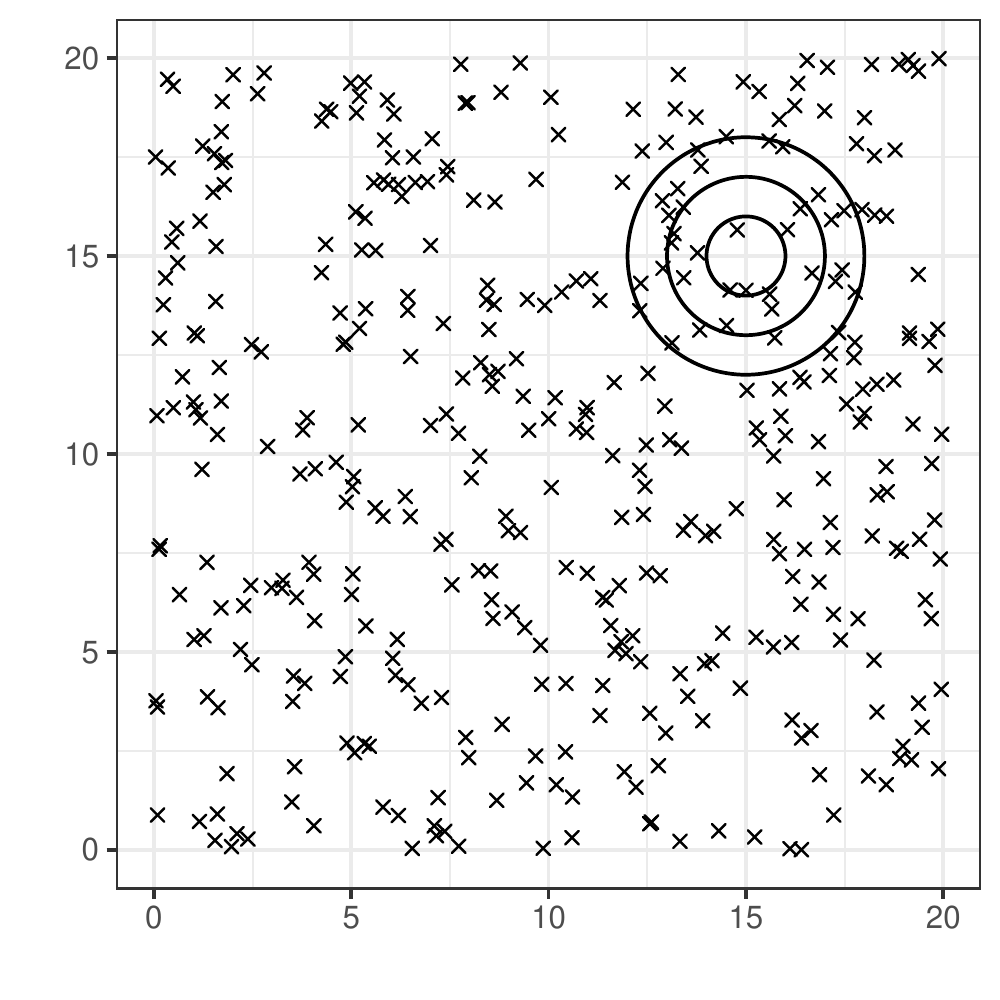}
    \end{minipage}
    \hfill
    \begin{minipage}[t]{0.42\textwidth}
        \centering
        \includegraphics[width=\linewidth]{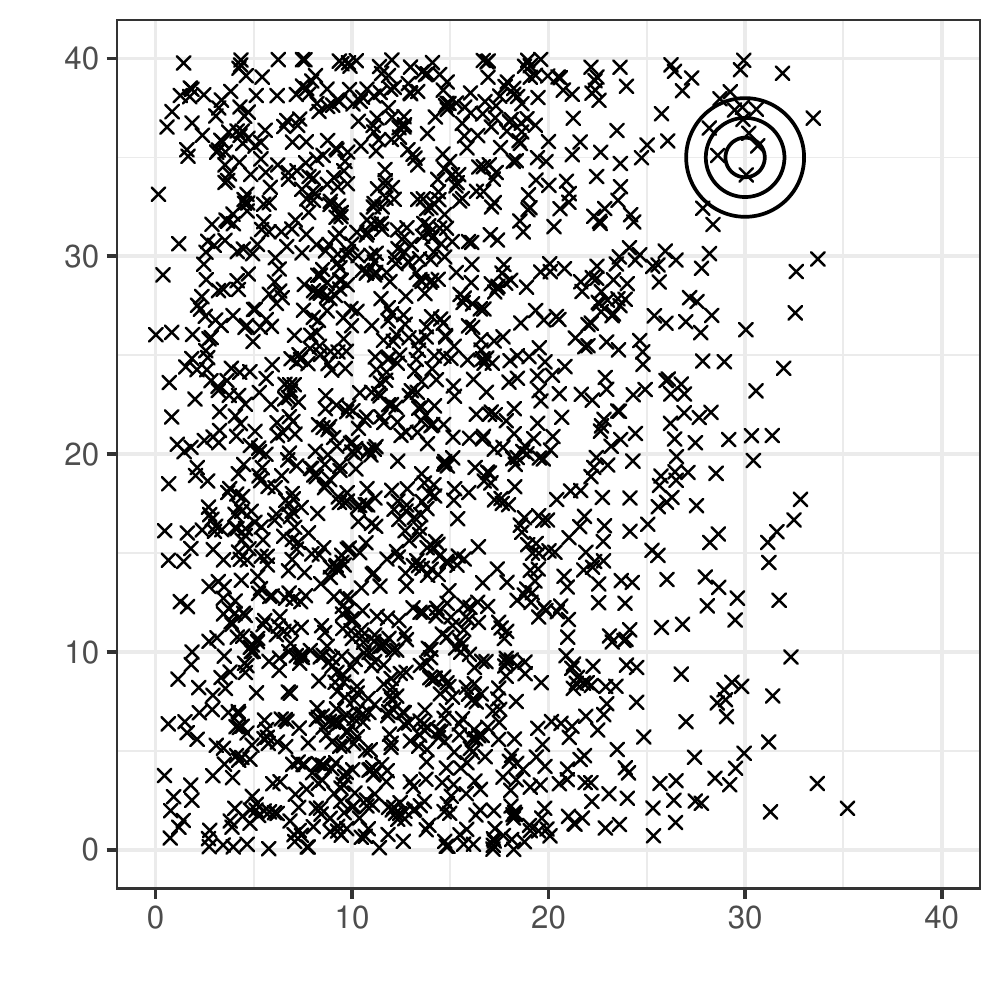}
    \end{minipage}
    \caption{Example sample locations for the uniform pattern for a domain of size $20 \times 20$ (left) and for the skew pattern for a domain of size $40 \times 40$ (right). The circles of radii 1, 2 and 3 represent the kernel function choices for SBSS.}
    \label{fig:sample_locs}
\end{figure}

\section{Simulations}\label{sec:sim}

In this part we validate the performance of our above introduced estimators by carrying out experiments on synthetic datasets. We use the statistical software R 3.6.1 (\cite{r_software}) with the help of the packages JADE (\cite{JADE_package}), SpatialBSS (\cite{SpatialBSS_package}) and RandomFields (\cite{RandomFields_package}).

We consider the case of $d=2$ and domains of the form $[0,l] \times [0,l]$ with $l \in \{10,20,30,40,50,60\}$ denoted as $l \times l$ where the sample locations are either following a uniform or a skewed design. The distributions for the entries of sample locations $\s = (s_1,s_2)^\top$ are $s_1 \sim U(0,1)$ and $s_2 \sim U(0,1)$ for the uniform setting and $s_1 \sim \beta(2,4)$ and $s_2 \sim U(0,1)$ for the skewed setting, where $U$ and $\beta$ denote the uniform and beta distributions, respectively. A set of sample locations for a given domain is formed by sampling $l^2$ iid samples of the former distributions which are then multiplied by $l$. Figure~\ref{fig:sample_locs} depicts one $20 \times 20$ example for the uniform pattern on the left panel and one $40 \times 40$ skewed pattern example on the right panel. The considered random field models which are simulated on the sample location patterns are discussed in the subsequent chapters.

\begin{figure}[t]
	\centering
  \includegraphics[width=0.5\linewidth]{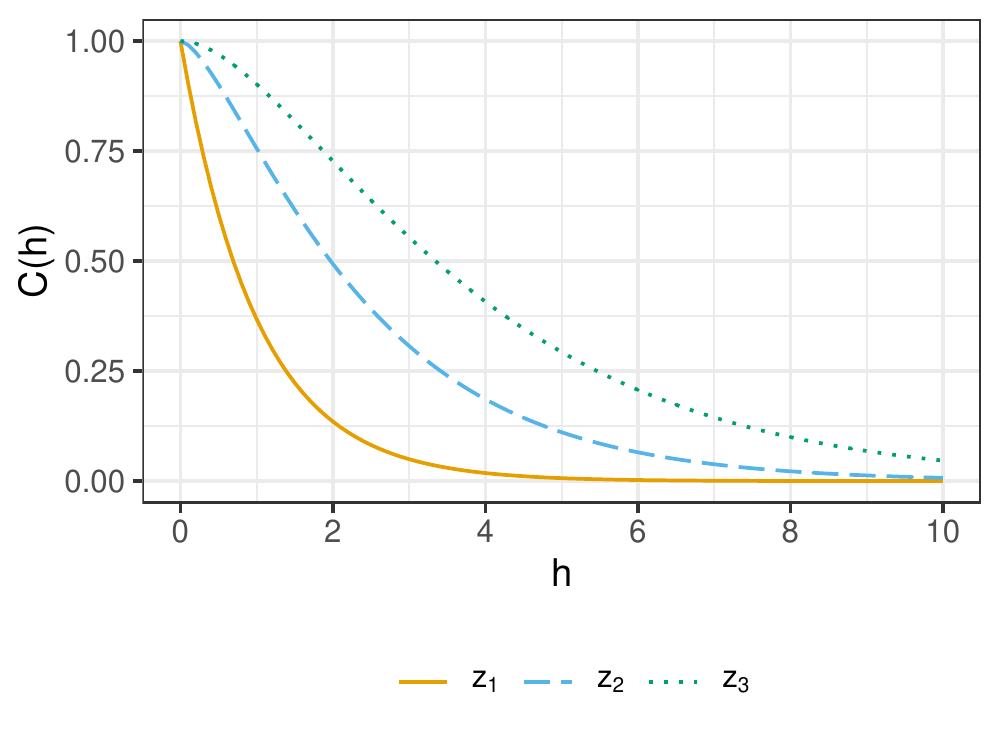}
  \caption{M\'atern covariance functions for the entries of the latent random field $\z$. The parameters are chosen to be $(\sigma, \nu, \phi) \in \{ (1.0, 0.5, 1.0), (1.0, 0.9, 1.7), (1.0, 1.3, 2.2)\}$.}
  \label{fig:covs}
\end{figure}

For a given random field we estimate the unmixing matrix by the following five BSS methods. The first two estimators are SBSS with $\LCov$ matrices from Definition~\ref{def:sbss_sd} using a ring kernel with the parameter $r=1$ (LCov Ball), and Definition~\ref{def:sbss_jd} using three ring kernels with parameters $(r_i,r_o) \in \{ (0,1), (1,2), (2,3) \}$ (LCov Ring). Furthermore, we use two SBSS estimators with $\LDiff$ matrices, namely the one from Definition~\ref{def:sbss_ldiff_sd_1} using a ring kernel with the parameter $r=1$ (LDiff Ball), and the one from Definition~\ref{def:sbss_ldiff_sd_2} where $f_1$ is the ring kernel function with $(r_i,r_o) = (0,1)$ and $f_2$ is also a ring kernel function with $(r_i,r_o) = (1,2)$ (wLDiff Ring). Lastly, we use the well known forth order blind identification (FOBI) algorithm \cite{Cardoso1989} which is an ICA method designed for iid data and therefore does not utilize spatial dependence information at all.

We evaluate the quality of the unmixing matrix estimation by computing the minimum distance index (MDI) \cite{IlmonenEtAl2010,LietzenVirtaNordhausenIlmonen2020} which is defined by 
\[
	\text{MD}(\hat{\bs W} \bs A) = \frac{1}{\sqrt{p-1}} \inf_{\bs M \in \mathcal{M}} \| \bs M \hat{\bs W} \bs A - \bs I_p \|_F .
\]
Here, $\hat{\W}$ is the unmixing matrix estimate, and $\mathcal{M}$ is the set of all matrices of the form $\bs P \bs D \bs S$, where $\bs P$ is a permutation matrix, $\bs D$ is a diagonal matrix with positive diagonal entries and $\bs S$ is a sign change matrix. As discussed above, $\hat{\W} \A$ should equal $\bs I_p$ up to scale, sign and order which are exactly the indeterminacies captured by the set $\mathcal{M}$. Loosely, the MDI measures the deviation of $\hat{\W} \A$ from $\bs I_p$ in respect of the indeterminacies, where it takes values between $0$, indicating a perfect estimate, and $1$, indicating a poor estimate. Note that the MDI only depends on $\hat{\W} \A$ which is always equal to the unmixing matrix evaluated for the case of $\A = \bs I_p$ for affine equivariant BSS methods, see for example \cite{bachoc2018spatial} for details. Hence, the results are equal independently of the specific form of $\A$, which favors $\A = \bs I_p$ as a convenient choice for simulations. In the following we present results for two different random field models.

\begin{figure}[t]
	\centering
  \includegraphics[width=0.8\linewidth]{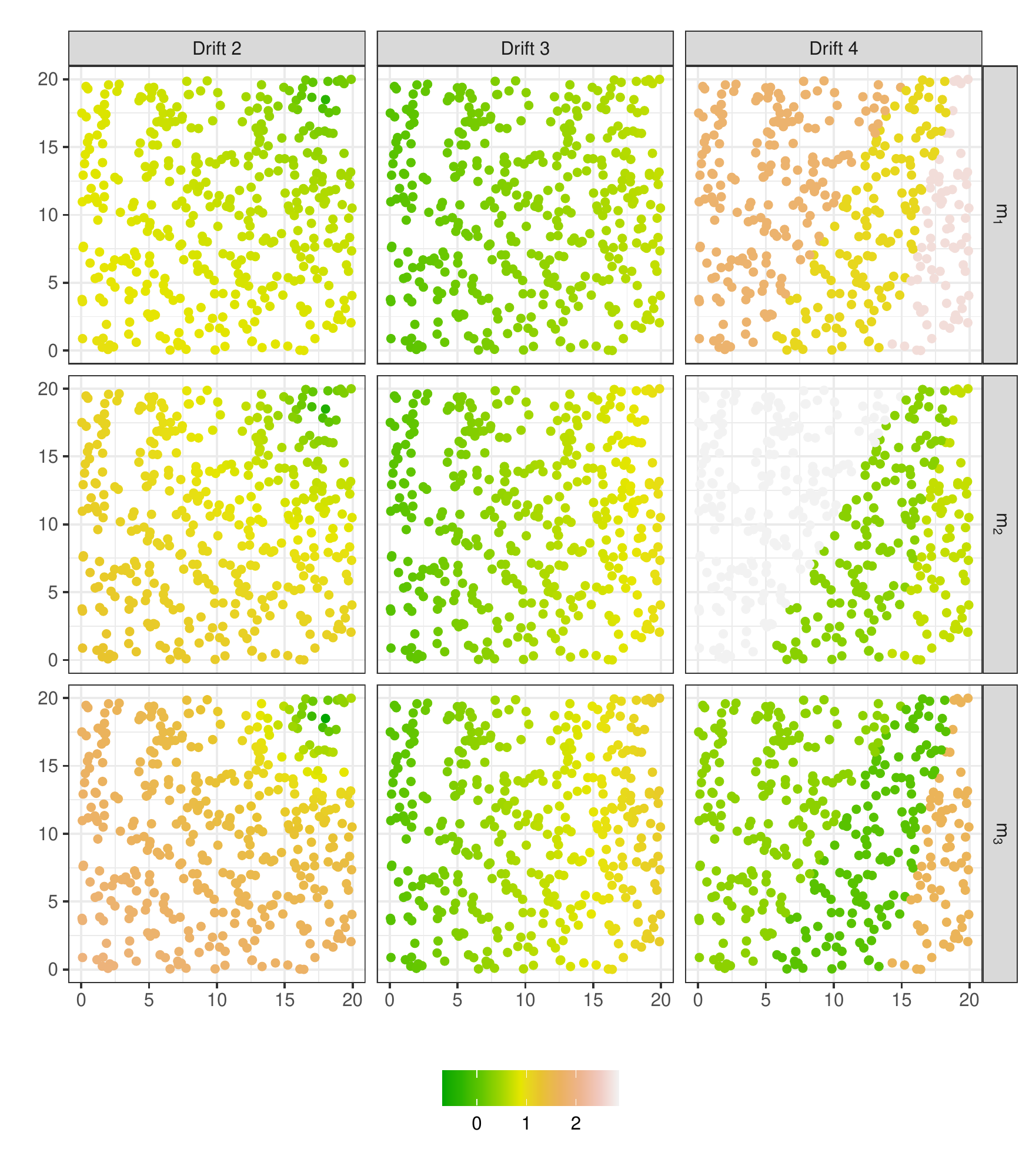}
  \caption{Drifts for one simulation replicate of drift Models 2-4 on a domain of size $20 \times 20$.}
  \label{fig:drift_example}
\end{figure}

\subsection{Weakly Stationary Latent Field with External Drift}

In this simulation, we consider the random field model of Equation~\eqref{eq:sbss} where we chose $p=3$ and $\A = \bs I_3$. The latent random field is formed by three centered Gaussian random fields, where the second order dependence is determined by the well-known second order stationary M\'atern covariance function (see for example \cite{ChilesetAl1999}) defined by
\[
C(h;\sigma^2, \nu, \phi) = \frac{\sigma ^ 2}{2 ^ {\nu - 1} \Gamma (\nu)} \left( \frac{h}{\phi} \right) ^ \nu  K_\nu \left( \frac{h}{\phi} \right), ~
h = \| \s - \s' \|.
\]
Here, $\sigma^2, \nu$ and $\phi$ are the variance, shape and range parameters respectively. $K_\nu$ denotes the modified Bessel function of second kind. The parameters are chosen to be $(\sigma^2, \nu, \phi) \in \{ (1.0, 0.5, 1.0), (1.0, 0.9, 1.7), (1.0, 1.3, 2.2)\}$ which are illustrated in Figure~\ref{fig:covs}. For the $3$-dimensional drift function $\m (\s) = (m_1(\s),m_2(\s),m_3(\s))^\top$ we consider the following four settings.

\paragraph{Drift 1:} For this case we choose $\m (\s) = \bs 0$ for all sample locations. Therefore, this model reduces to the constant drift case which was examined in detail in \cite{bachoc2018spatial,nordhausen2015blind}.

\paragraph{Drift 2:} This drift is radial-symmetric in its nature, specifically, it is formed by $m_i(\s) = c_i \log(\| \s_0 - \s \|)$ for $i=1,2,3$ where $\s_0$ is one artificial sample location sampled uniformly inside the domain at hand. Hence, for a domain of size $l \times l$ the entries of the artificial location $\s_0 = (s_{01},s_{02})^\top$ are drawn from $s_{01},s_{02} \sim U(0,l)$. The constants are chosen to be $(c_1, c_2, c_3) = (0.3, 0.4, 0.6)$. Theoretically, the drift $m_i$ ranges between $(- \infty,  2.6 c_i]$ for a $10 \times 10$ domain up to $(- \infty, 4.5 c_i]$ for a $60 \times 60$ domain.

\paragraph{Drift 3:} Here, we consider a linear drift in the first direction of the sample locations. For $i=1,2,3$, $m_i(\s)$ equals $c_i s_1 / s_1^*$ where the constants are $(c_1, c_2, c_3) = (0.7, 1, 1.2)$. $s_1$ is the first entry of the sample locations $\s$ and $s_1^*$ is the maximum value of all first entries for the given set of sample locations. Therefore, the trend ranges between $(0, c_i]$ for all different domain sizes.

\paragraph{Drift 4:} In the same fashion as in Drift 2 we sample three artificial sample locations inside the domain at hand. These artificial locations define three cluster centers, a point belongs to the cluster where the Euclidean distance to the cluster center is minimal. For each cluster $j=1,2,3$ the corresponding drift $m_i^j$ is a sample from the uniform distribution $U(0,3)$. Consequently, the drift for this model lies in the interval $[0,3]$. Overall, this setting is constituted by an observable random field $\x$ that is weakly stationary in each cluster of sample locations, which often denoted as a block-stationary model.

One example for the different drift settings described above is depicted in Figure~\ref{fig:drift_example}. Note that the on-sight variance for each entry of $\z$ equals one, moreover, the latent field is Gaussian distributed. Therefore, over 99\% of the values of $\z$ are in the interval $[-3,3]$ which highlights the impact of the ranges of the different drift settings. 

\begin{figure}[t]
	\centering
  \includegraphics[width=0.75\linewidth]{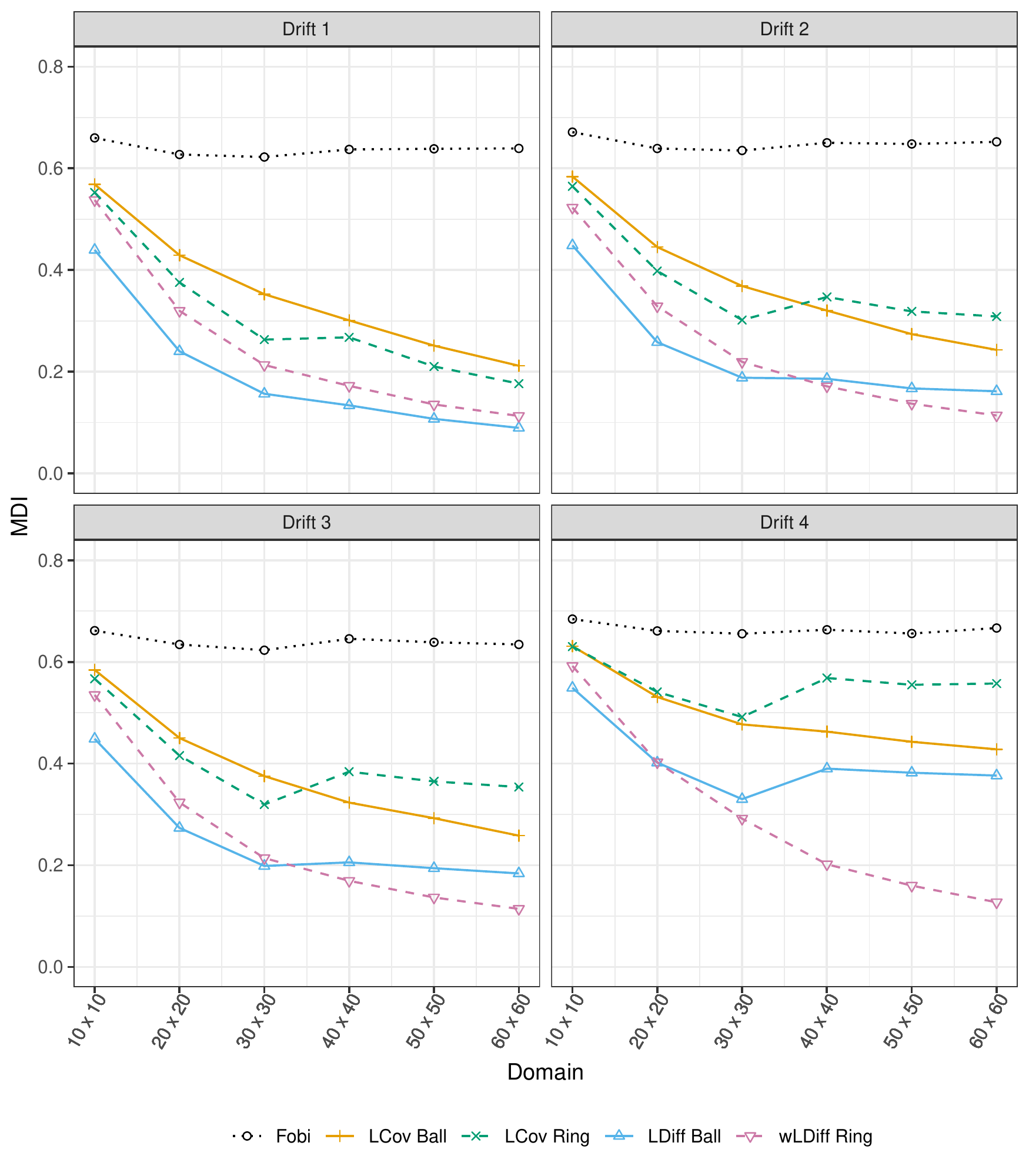}
  \caption{Average MDI based on 2000 iterations for the uniform sample location pattern where the observed random field is formed by a weakly stationary latent field and four different additive drift functions.}
  \label{fig:md_unif}
\end{figure}

\begin{figure}[t]
	\centering
  \includegraphics[width=0.75\linewidth]{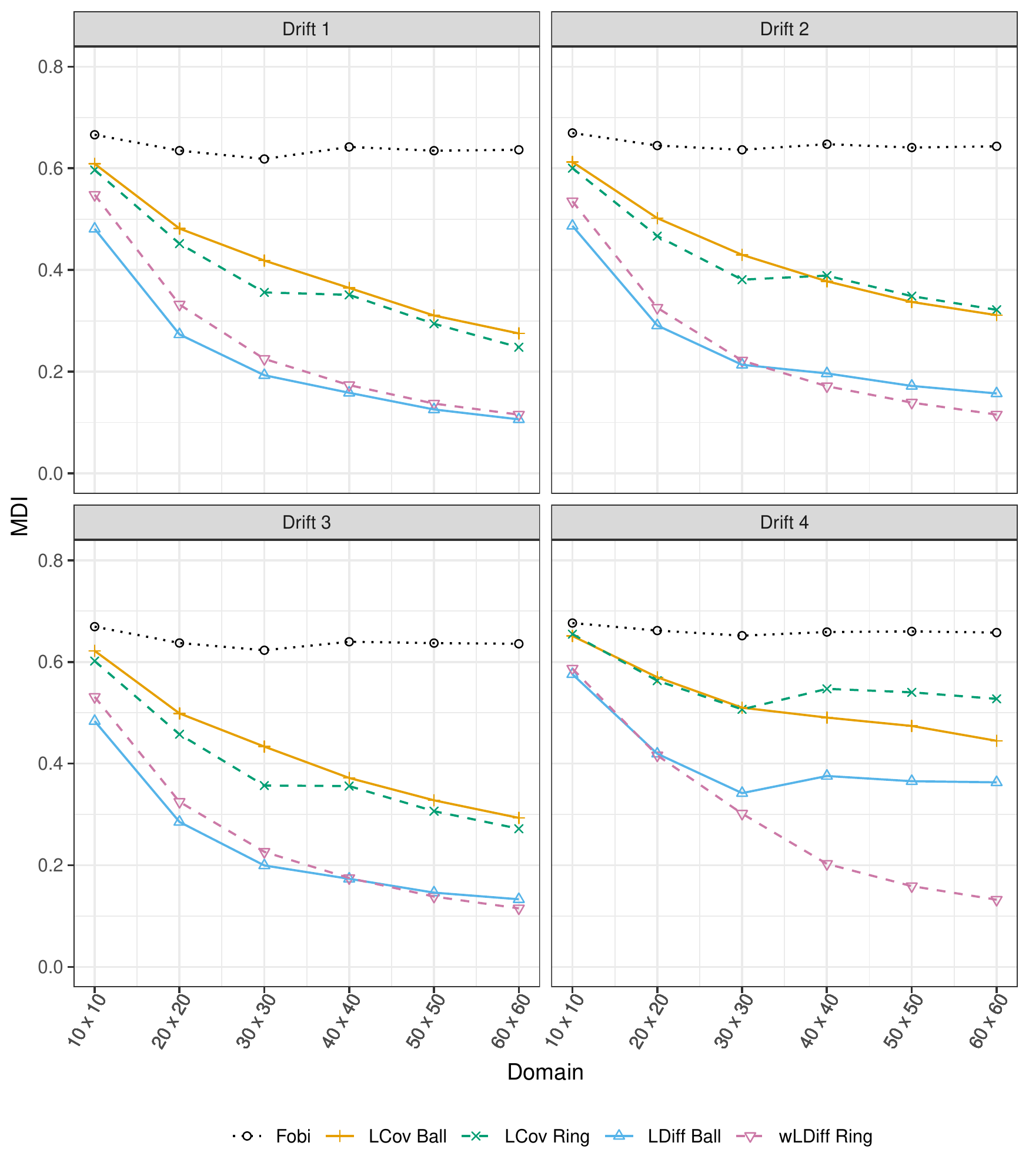}
  \caption{Average MDI based on 2000 iterations for the skewed sample location pattern where the observed random field is formed by a weakly stationary latent field and four different additive drift functions.}
  \label{fig:md_skew}
\end{figure}

The average MDI values based on 2000 repetitions for the uniform coordinate pattern are presented in Figure~\ref{fig:md_unif}. Overall, for all considered drift settings the SBSS methods that are based on local difference matrices show much better performance. This is especially surprising for Drift 1 as it follows the original SBSS model. The adapted whitening step seems to be of minor influence with Drift 4 being an exception, in this setting the performance of $\LDiff$ Ball seems to stay constant even if the sample size increases. Similarly, the performance difference between SBSS that simultaneously or jointly diagonalize $\LCov$ matrices is not as significant. Figure~\ref{fig:md_skew} depicts the average MDI for the skewed sample location pattern. The qualitative meaning is very similar to the uniform setting with two minor exceptions. Firstly, the overall MDI is slightly higher for all settings and estimators, this might be explained by the fact that sample locations are only dense on the left part of the domain. Therefore, the effective sample size decreases when local covariance or local difference matrices are estimated, as certain sample locations do not have neighbors that are captured by the spatial kernel functions. Secondly, SBSS methods that are based on $\LDiff$ matrices do not suffer such a significant reduction of performance as the ones using $\LCov$ matrices.

\subsection{Intrinsic Stationary Latent Field with Constant Drift}

\begin{figure}[t]
	\centering
  \includegraphics[width=0.8\linewidth]{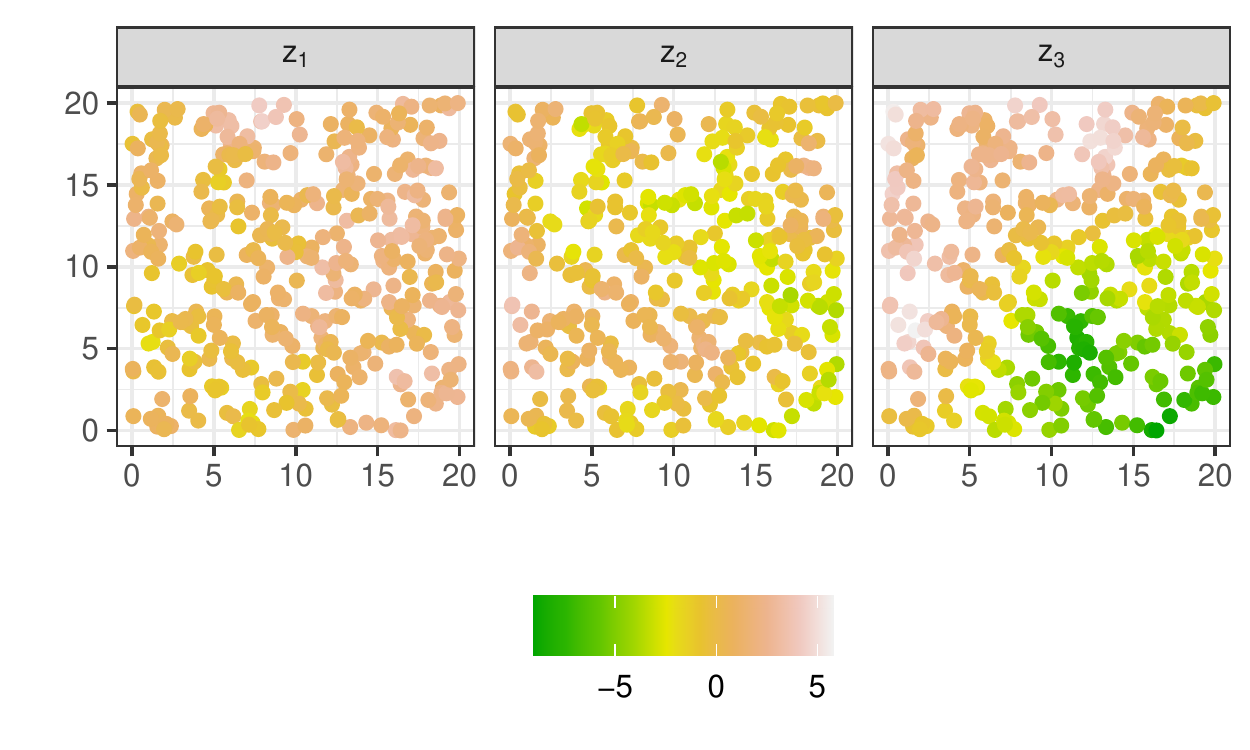}
  \caption{One example for the latent random field which follows the fractional Brownian field model on a $20 \times 20$ domain with uniform sample location pattern. The Hurst parameters equal 0.3, 0.5 and 0.8 for $z_1$, $z_2$ and $z_3$ respectively.}
  \label{fig:intr_example}
\end{figure}

In this simulation we again consider the SBSS model of Equation~\eqref{eq:sbss} where $p=3$, $\A= \bs I_3$ and $\m = \bs 0$. In contrast to Definition~\ref{def:sbss_model} the latent random field is now formed by uncorrelated intrinsic stationary random fields, which is an example for the outline in Section~\ref{sec:intrinsic_stat_sbss}. Specifically, we chose the second order dependence of the entries of $\z$ to follow the fractional Brownian field covariance function (see for example \cite{FracBrown}) defined by
\[
	C(\s, \s';H) = \frac{1}{2} \left( \| \s \| ^{2H} + \| \s' \| ^{2H} - \| \s - \s' \| ^{2H} \right).
\]
In this equation, $H \in (0,1]$ is denoted as the Hurst parameter. The fractional Brownian field is a well-known intrinsic stationary but not weakly stationary random field model. We choose the Hurst parameters for the latent field to be $H \in \{0.3, 0.5, 0.8\}$, where Figure \ref{fig:intr_example} depicts one example.

\begin{figure}[t]
	\centering
  \includegraphics[width=0.75\linewidth]{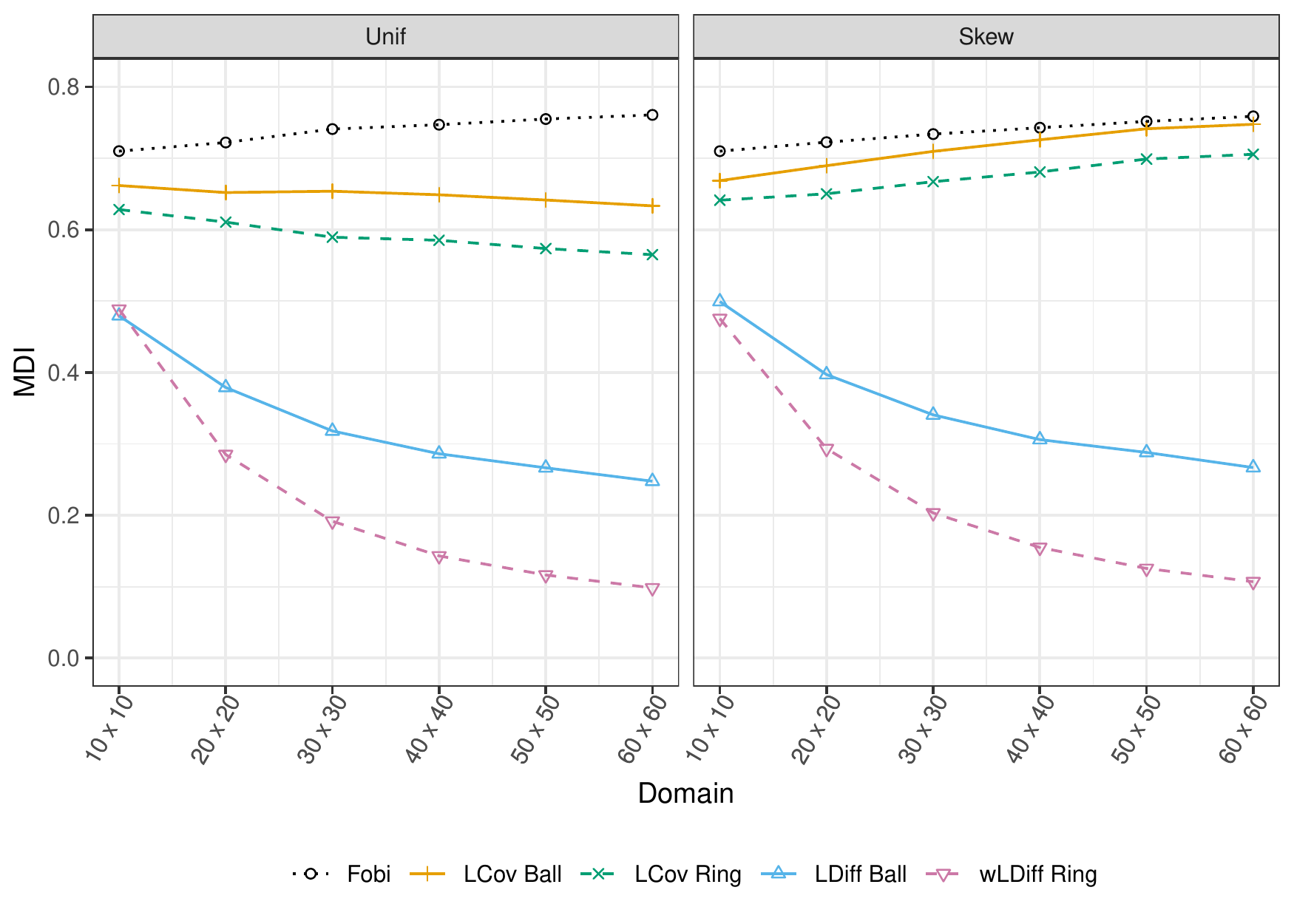}
  \caption{Average MDI based on 2000 repetitions for the simulation with intrinsic stationary latent fields.}
  \label{fig:md_brown}
\end{figure}

The average MDI values based on 2000 simulation iterations for the uniform and skewed sample location pattern are presented in Figure~\ref{fig:md_brown}. The SBSS methods that only rely on the use of $\LCov$ matrices show a very poor performance. This is expected as the spatial covariance of $\z$ depends on the actual sample location, which in turn leads to the fact that proper estimation of $\LCov$ matrices is impossible. In contrast, the SBSS methods relying on $\LDiff$ matrices still perform well, because, the variance of the difference processes of the random field at hand still depends only on the distance between sample locations. Clearly, the method with the adapted whitening step shows the overall best performance as also the whitening step is not corrupted by the intrinsic stationary nature of the observed random field $\x$.

\section{Data Example}\label{sec:real_data}

In the following we illustrate the usefulness of our proposed adaptation of SBSS on a dataset derived from the GEMAS project \cite{ReimannEtAl2014}, which was also considered in \cite{CAJG_paper}. The dataset is freely available in the R package robCompositions (\cite{robCompositions_package}) and it is formed by concentration measurements in mg/kg of 18 elements (Al, Ba, Ca, Cr, Fe, K, Mg, Mn, Na, Nb, P, Si, Sr, Ti, V, Y, Zn, Zr) at 2107 samples of agricultural soils in Europe, with locations given in latitude and longitude coordinates. 

As it is common practice in regional geochemistry we respect the relative information of the data by using principles of compositional data analysis \cite{comp_book}. Specifically, we carry out a three step analysis. Firstly, we transform the dataset into centered log-ratio (clr) coordinates. Clr coordinates are easy to interpret but have the disadvantage that each observation adds up to zero, which in turn results in the fact that for example the covariance matrix is not invertible, something which is needed in BSS. Therefore, we transform the data into isometric log-ratio (ilr) coordinates by using pivot coordinates, which is only an orthogonal transformation of the clr data, this orthogonal matrix is usually denoted as contrast matrix (\cite{robCompositions_paper}). As ilr coordinates are indeed of full rank, all SBSS methods can be carried out in that space and the so-called combined loadings matrix consists of the matrix product of the contrast matrix and the estimated unmixing matrix. Interpretations of the results are carried out with the combined loadings matrix in terms of the clr coordinates. This procedure is detailed in \cite{nordhausen2015blind}, where the original SBSS is applied in the context of regional chemistry. The ilr transformation reduces the dimension of the data from $p=18$ to $p=17$.

It was already shown in \cite{CAJG_paper} that the norms of the difference between the global average and the measurements on-site are not constant throughout the spatial domain, which suggests to use $\LDiff$ matrices in favor of $\LCov$ matrices. In contrast to \cite{CAJG_paper} where the method of Definition~\ref{def:sbss_ldiff_sd_1} is utilized, we analyze the data with the adapted whitening step seen in Definition~\ref{def:sbss_ldiff_sd_2}. Specifically, we chose $f_1$ and $f_2$ to be ring kernel functions with the parameters $(0^{\circ},2^{\circ})$ and $(2^{\circ},4^{\circ})$ respectively. 

\begin{figure}[t]
\centering
    \begin{minipage}[t]{0.45\textwidth}    
        \centering
        \includegraphics[width=\linewidth]{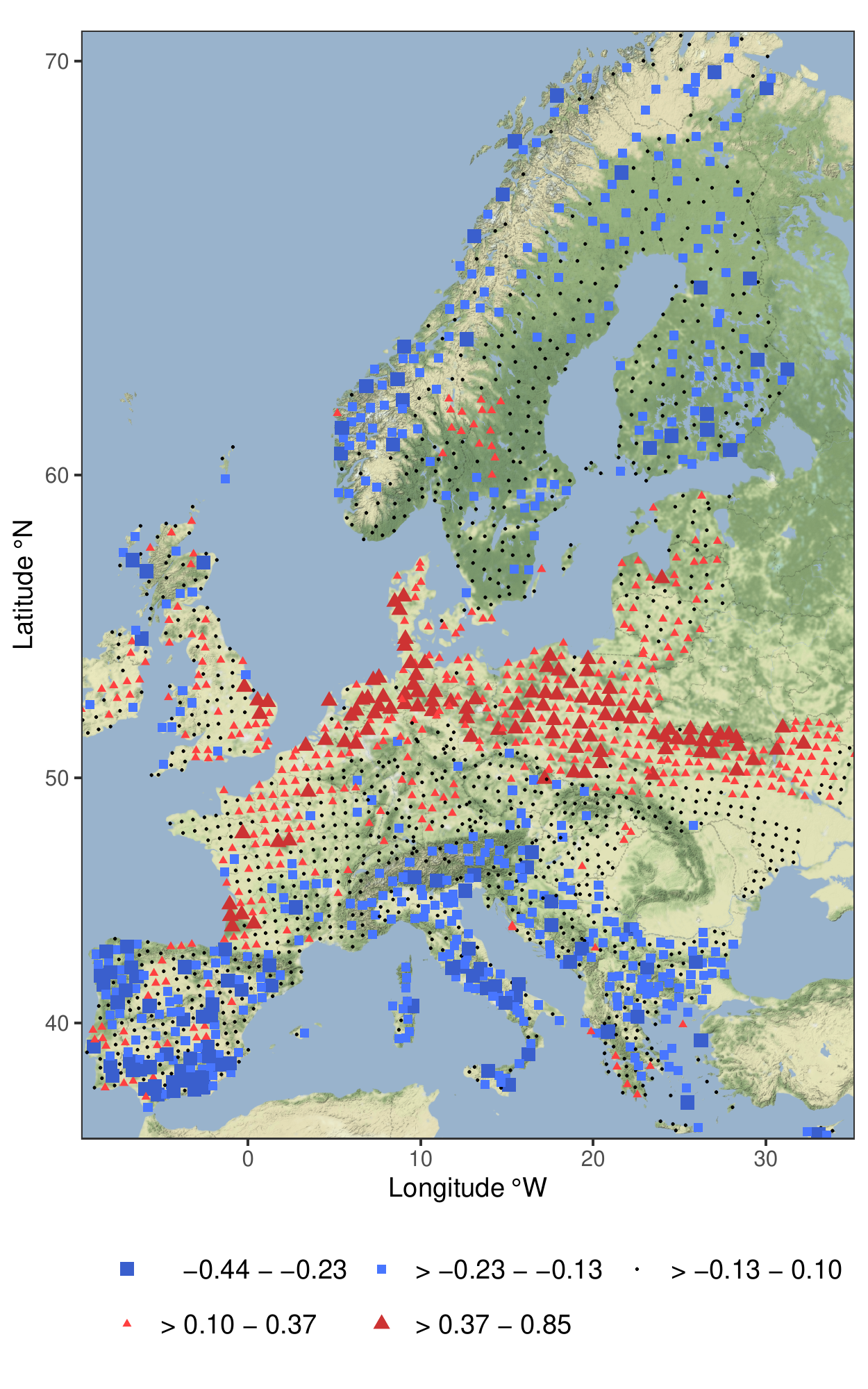}
    \end{minipage}
    \hfill
    \begin{minipage}[t]{0.45\textwidth}
        \centering
        \includegraphics[width=\linewidth]{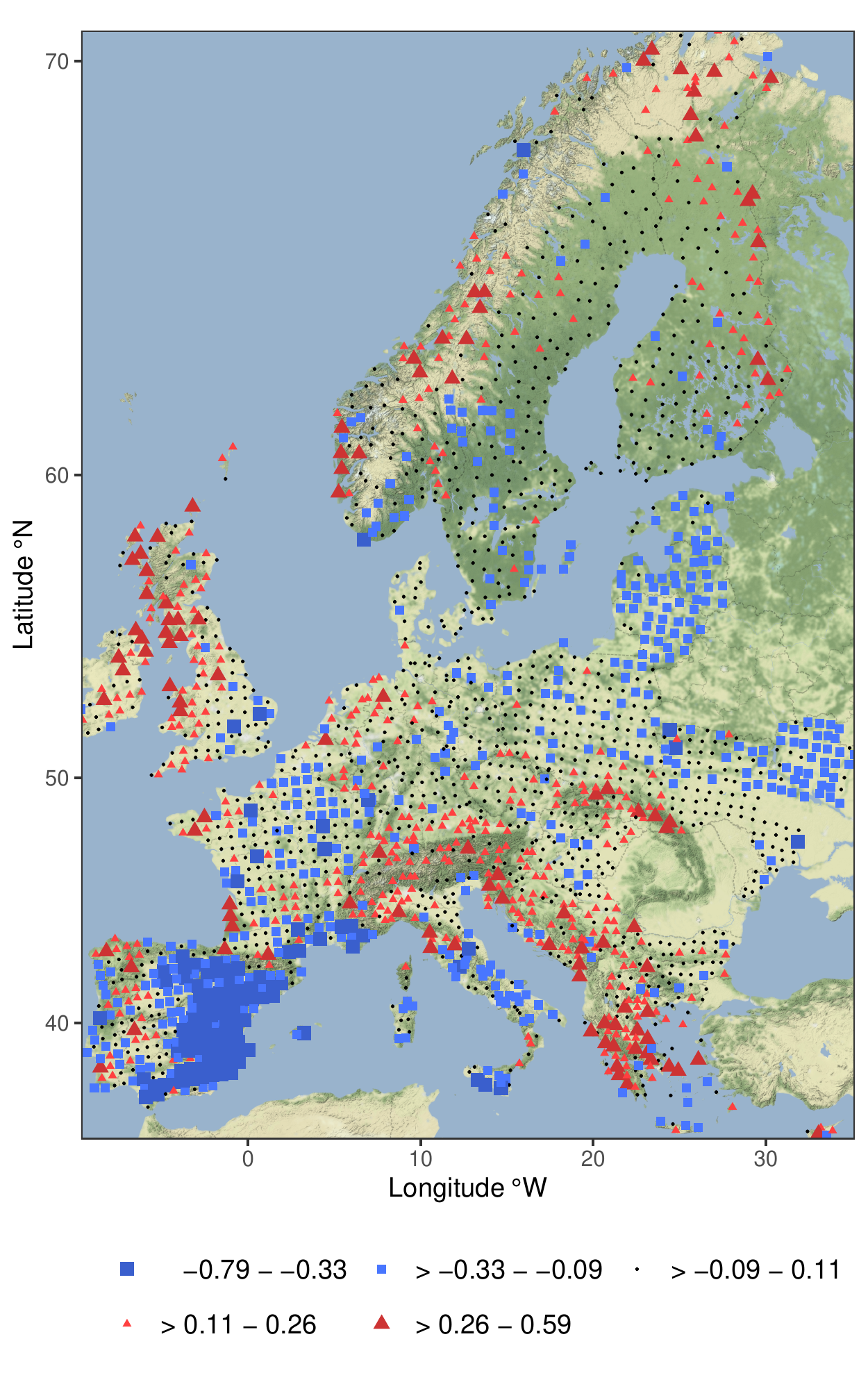}
    \end{minipage}
    \caption{First (left) and second (right) entry of the estimated latent random field with the method of Definition~\ref{def:sbss_ldiff_sd_2}. Where, $f_1$ and $f_2$ are ring kernel functions with parameters $(0^{\circ},2^{\circ})$ and $(2^{\circ},4^{\circ})$. Map tiles by Stamen Design, under CC BY 3.0. Data by OpenStreetMap, under ODbL.}
    \label{fig:gemas_ics}
\end{figure}

\begin{table}[t]
\centering
\caption{Combined loadings for the first ($z_1$) and second ($z_2$) entry of the latent random field.}
\begin{tabular}{lrrlrr}
  \toprule
 & $z_1$ & $z_2$ & &  $z_1$ & $z_2$ \\ 
  \midrule
  clr(Al) & \textbf{-0.30} & \textbf{0.20} &       clr(Nb) & 0.00 & -0.02 \\ 
  clr(Ba) & -0.06 & -0.02 &    clr(P) & 0.04 & 0.09 \\ 
  clr(Ca) & 0.02 & -0.08 &     clr(Si) & \textbf{0.17} & 0.01 \\ 
  clr(Cr) & 0.05 & 0.09 &      clr(Sr) & -0.03 & -0.06 \\ 
  clr(Fe) & -0.04 & \textbf{-0.23} &    clr(Ti) & -0.05 & -0.07 \\ 
  clr(K) & 0.11 & \textbf{-0.20} &     clr(V) & 0.11 & \textbf{0.14} \\ 
  clr(Mg) & -0.08 & 0.03 &     clr(Y) & -0.10 & -0.02 \\ 
  clr(Mn) & -0.01 & 0.02 &     clr(Zn) & 0.06 & 0.06 \\ 
  clr(Na) & 0.01 & 0.12 &      clr(Zr) & 0.08 & -0.04 \\ 
   \bottomrule
\end{tabular}
\label{tab:loadings_ics}
\end{table}

Figure~\ref{fig:gemas_ics} presents the first two entries of the estimated latent field and Table~\ref{tab:loadings_ics} shows the corresponding first two rows of the combined loadings matrix. The first entry is mainly driven by Aluminum and Silicium based on the high absolute values of the combined loadings. Therefore, in the glacial sediments of northern central Europe this indicates that Silicium is a dominant element, whereas Aluminum is more dominant in the Southern areas, which is confirmed by the original clr maps. For the second component it is interesting to see that the combined loadings for Aluminum and Vanadium as well as Iron and Potassium show roughly the same absolute values with different signs. 
Accordingly, Iron and Potassium are dominating in the geochemical composition of
the soils in southern and eastern Spain, where the soils are formed differently than
in the remaining part of the country and in Portugal. This dominance is also visible in southern Italy, in the Baltic countries and in parts of the Ukraine. In contrast, dominance of Aluminum-Vanadium is observed in soils along the Alps, in parts of the Balkan Peninsula and the Carpathian Mountains, but also in the western part of the UK and in Norway.
A thorough interpretation of the results might be carried out by domains experts.

\section{Discussion and Conclusions}\label{sec:conclusion}

In this paper we recalled SBSS, which is an unsupervised statistical tool that finds uncorrelated latent fields given a multivariate observed random field. This methodology offers one way to deal with multivariate spatial data in such a way that univariate methods can be used, a very appealing approach as multivariate spatial data is generally demanding to model. For the original SBSS method this useful feature was already investigated for interpolation tasks in \cite{kriging_paper}. However, the original SBSS methods rely on the strong assumption that the drift is constant for the whole considered spatial domain, which is a very restrictive assumption. We introduced a new scatter matrix which measures second order spatial dependence based on differences, and argued that replacing local covariance matrices with local difference matrices yields great advantages when a drift is present in the data. We also adapted the whitening step of the SBSS methods by such a replacement, confirmed our outline in simulations and showed the usefulness of our new adaptations on a geochemical dataset derived from the GEMAS project.

The original motivation for the adapted whitening step seen in \cite{BelouchraniCichocki2000} is in the context of BSS for time series data with additive white noise. Such a model can also be considered for the spatial case, in which the drift $\m$ would be replaced by a centered $p$-variate white noise process $\bs n(\s)$ with $\Cov(\bs n(\s)) = \M$ in Equation~\eqref{eq:sbss}. This might be viewed as an external nugget effect. In such a case considering $f_0$, ball or Gauss spatial kernel functions $\LCov_{f}(\x)$ equals $\A \LCov_{f}(\z) \A ^ \top + \M$, but for a ring kernel function the additive term $\M$ would disappear. Therefore, proper estimation of $\W$ would be carried out with an adapted whitening step were only ring kernel functions are used. The problem here would be that local covariance functions with ring kernels are not necessarily positive definite, therefore a more sophisticated algorithm from \cite{BelouchraniCichocki2000} has to be adapted, which we plan for future research. Note that for this case $\LDiff$ matrices would be useless as they always carry on-sight covariance terms independent of the used spatial kernel function.

\section*{Acknowledgement}

The work of CM and KN was supported by the Austrian Science Fund P31881-N32.


\end{document}